%% file: main.tex
\documentclass[sigconf, nonacm]{acmart}

\AtBeginDocument{%
  }

\copyrightyear{2025}
\acmYear{2025}
\setcctype{cc-by}
\usepackage{colortbl}
\usepackage{tikz}
\usepackage{array}
\usepackage{multirow}
\usepackage{fontawesome}
\usepackage{array}
\usepackage{subcaption}
\newcolumntype{C}[1]{>{\centering}m{#1}}
\usepackage{CJKutf8}

\begin{document}

% \title[WebFAQ]{WebFAQ: Extracting Question and Answer pairs from the web}
\title[WebFAQ 2.0: A Multilingual QA Dataset with Mined Hard Negatives for Dense Retrieval]{WebFAQ 2.0: A Multilingual QA Dataset\\with Mined Hard Negatives for Dense Retrieval}

\author{Michael Dinzinger}
\email{michael.dinzinger@uni-passau.de}
% \authornote{Corresponding author}
\orcid{0009-0003-1747-5643}
\affiliation{%
  % \department{Chair of Data Science}
  \institution{University of Passau}
  \streetaddress{Innstraße 33}
  \city{Passau}
  \country{Germany}
  \postcode{94032}
}

\author{Laura Caspari}
\email{laura.caspari@uni-passau.de}
\orcid{0009-0002-6670-3211}
\affiliation{%
  % \department{Chair of Data Science}
  \institution{University of Passau}
  \streetaddress{Innstraße 33}
  \city{Passau}
  \country{Germany}
  \postcode{94032}
}

% \author{Kanishka Ghosh Dastidar}
% \email{kanishka.ghoshdastidar@uni-passau.de}
% \orcid{0000-0003-4171-0597}
% \affiliation{%
%   % \department{Chair of Data Science}
%   \institution{University of Passau}
%   \streetaddress{Innstraße 33}
%   \city{Passau}
%   % \state{Ohio}
%   \country{Germany}
%   \postcode{94032}
% }

\author{Ali Salman}
\email{salman05@ads.uni-passau.de}
% \orcid{}
\affiliation{%
  % \department{Chair of Data Science}
  \institution{University of Passau}
  \streetaddress{Innstraße 33}
  \city{Passau}
  \country{Germany}
  \postcode{94032}
}

\author{Irvin Topi}
\email{topi01@ads.uni-passau.de}
\orcid{0009-0001-8298-5039}
\affiliation{%
  % \department{Chair of Data Science}
  \institution{University of Passau}
  \streetaddress{Innstraße 33}
  \city{Passau}
  \country{Germany}
  \postcode{94032}
}

\author{Jelena Mitrovi\'{c}}
\email{jelena.mitrovic@uni-passau.de}
\orcid{0000-0003-3220-8749}
\affiliation{%
  % \department{Chair of Data Science}
  \institution{University of Passau}
  \streetaddress{Innstraße 33}
  \city{Passau}
  \country{Germany}
  \postcode{94032}
}

\author{Michael Granitzer}
\email{michael.granitzer@uni-passau.de}
\orcid{0000-0003-3566-5507}
\affiliation{%
  % \department{Chair of Data Science}
  \institution{University of Passau, and IT:U Austria}
  \streetaddress{Innstraße 33}
  \city{Passau}
  \country{Germany}
  \postcode{94032}
}

% \affiliation{%

% \affiliation{%

\renewcommand{\shortauthors}{Dinzinger et al.}

\begin{abstract}
% We introduce WebFAQ 2.0, a new version of the WebFAQ dataset, containing 180 million FAQ-based natural question-answer pairs across 90 languages. Compared to the previous version, it significantly expands multilingual coverage and the number of bilingual aligned QA pairs, making it the largest FAQ-based resource. In response to community feedback, we also release a hard negatives dataset for training dense retrievers. These hard negatives were mined using a two-stage retrieval pipeline and filtered to avoid false negatives. Experiments show that dense retrieval models fine-tuned with these hard negatives outperform models trained with randomly sampled negatives. WebFAQ 2.0 thus provides a valuable resource for multilingual and cross-lingual IR.
We introduce WebFAQ 2.0, a new version of the WebFAQ dataset, containing 198 million FAQ-based natural question-answer pairs across 108 languages. Compared to the previous version, it significantly expands multilingual coverage and the number of bilingual aligned QA pairs to over 14.3M, making it the largest FAQ-based resource. Unlike the original release, WebFAQ 2.0 uses a novel data collection strategy that directly crawls and extracts relevant web content, resulting in a substantially more diverse and multilingual dataset with richer context through page titles and descriptions.
In response to community feedback, we also release a hard negatives dataset for training dense retrievers, with 1.25M queries across 20 languages. These hard negatives were mined using a two-stage retrieval pipeline and include cross-encoder scores for 200 negatives per query. We further show how this resource enables two primary fine-tuning strategies for dense retrievers: Contrastive Learning with MultipleNegativesRanking loss, and Knowledge Distillation with MarginMSE loss.

WebFAQ 2.0 is not a static resource but part of a long-term effort. Since late 2025, structured FAQs are being regularly released through the Open Web Index, enabling continuous expansion and refinement. We publish the datasets and training scripts to facilitate further research in multilingual and cross-lingual IR.
The dataset itself and all related resources are publicly available on GitHub\footnote{\url{https://github.com/padas-lab-de/webfaq}} and HuggingFace.\footnote{\url{https://huggingface.co/michaeldinzinger/webfaq-v2}}
\end{abstract}




% \received{05 February 2024}
% \received[revised]{-}
% \received[accepted]{-}

\maketitle

\section{Introduction}

\input{introduction}

\section{Related work}

\input{related_work}

\section{WebFAQ 2.0 Dataset Collection}

\input{data_collection}

\section{Hard Negatives Dataset Collection}

\input{hn_collection}

\section{Evaluation} \label{evaluation}

\input{evaluation}

% \section{Bilingual datasets}

% \input{parallel_corpus}

% \section{Discussion}

% \input{limitations}

\section{Conclusion}

\input{conclusion}

\begin{acks}
\begin{minipage}{0.30\linewidth}
\includegraphics[width=0.95\textwidth]{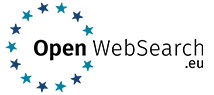}
\end{minipage}
\begin{minipage}{0.65\linewidth}
\href{https://doi.org/10.3030/101070014}{This work has received funding from the European Union's Horizon Europe research and innovation program under grant agreement No. 101070014 (OpenWebSearch.EU)}.
\end{minipage}
\end{acks}

\bibliographystyle{ACM-Reference-Format}
\bibliography{sources}

% \onecolumn

% \appendix

% \section{Appendix: Full Results} \label{appendix}
% The below tables list the NDCG@10 in \% results across models for all 20 WebFAQ retrieval corpora, the 18 languages included in MIRCAL (Hard Negatives) and the 11 languages of Mr. Tydi. The top three rows show the performance of three state-of-the-art dense embedding models: GTE-Multilingual-Base (mGTE), Multilingual-E5-Large-Instruct (mE5) and Jina (v3). The bottom four rows of each table compare the results of BM25, a popular traditional sparse embedding model, which we use as a baseline, to our pretrained XML-RoBERTa models. Here, \textbf{Base} refers to the model with in-domain pretraining on MS MARCO, while \textbf{FT} refers to the model that was additionally fine-tuned on WebFAQ data. \textbf{Hybrid} combines BM25 and FT as described in Section~\ref{model_fine_tuning}. Bold font indicates top values w.r.t. the first 3 and last 4 rows.

% \input{full_results_tables}

\end{document}

%% file: introduction.tex
Developing robust multilingual retrieval systems requires large-scale, diverse, and multilingual datasets, yet such resources remain limited in scope and scale. Frequently asked questions (FAQs), published as structured markup across websites worldwide, offer a rich and underutilized source of naturally occurring question-answer pairs. The datasets CCQA~\cite{Huber2022}, PAQ~\cite{Lewis2021} as well as the first version of WebFAQ~\cite{Dinzinger2025}, introduced last year, have demonstrated the potential of leveraging this data to improve multilingual dense retrievers. However, the original WebFAQ is constrained by data coverage, primarily due to its reliance on a limited set of preprocessed structured data dumps made available by the Web Data Commons project\footnote{\url{https://webdatacommons.org/}}, and it lacked explicit support for training with hard negatives.

\begin{figure}[t]
  \centering
  \includegraphics[width=0.95\linewidth]{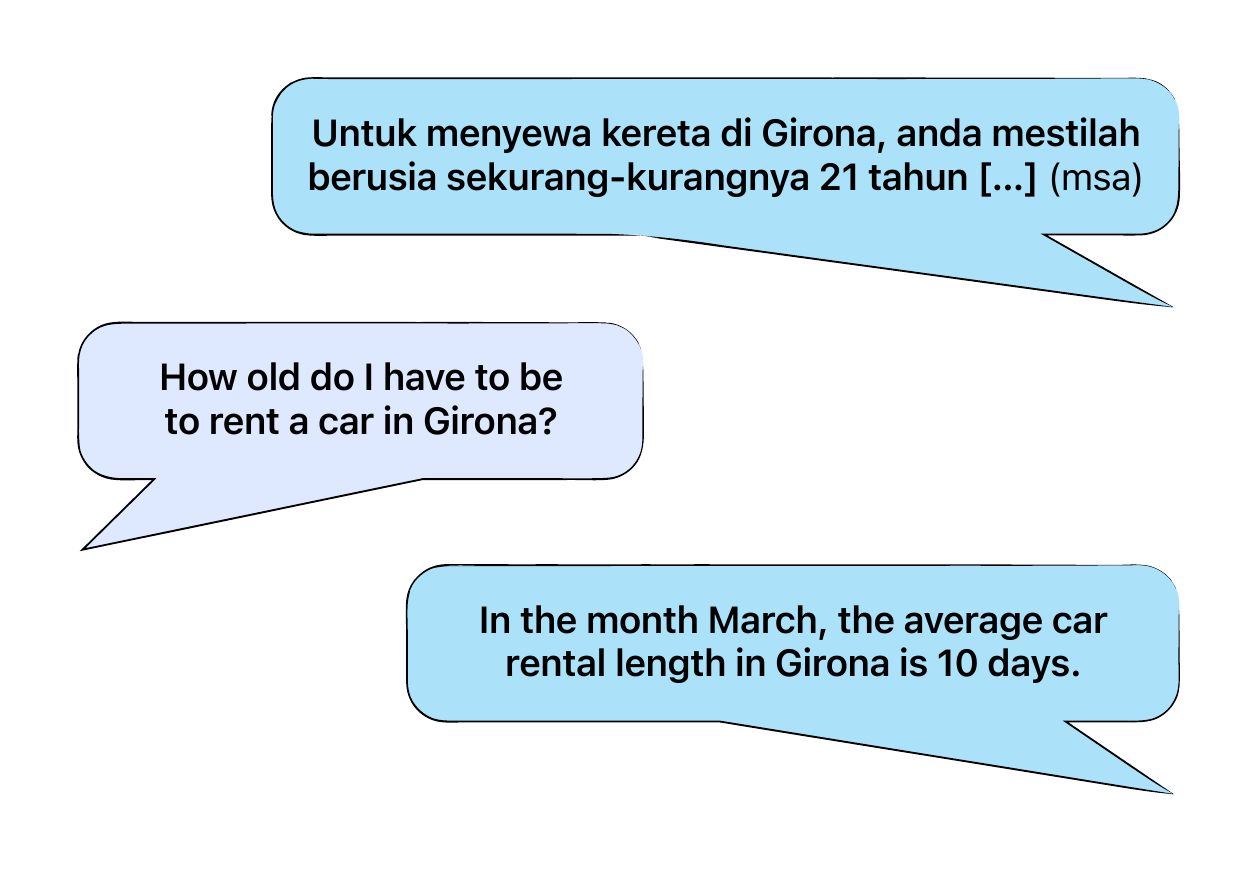}
  \vspace{-0.5cm}
  \caption{WebFAQ is a massive, publicly available repository of natural question-answer pairs. The second version focuses on QAs aligned across languages and hard negative mining.}
  \label{fig:nDCG_drop_jina}
  \Description{TODO}
  \vspace{-0.2cm}
\end{figure}

In this paper, we present WebFAQ 2.0, which introduces two key advancements. First, the dataset now includes 198 million QA pairs across 104 languages, representing more than double the size and vastly increased language diversity compared to the original release. This expansion was enabled by a fundamentally revised data collection strategy: instead of relying solely on structured data extractions, we mined URLs that potentially contain structured FAQs directly from Common Crawl and use the OWLer crawling tool~\cite{Dinzinger2024} to retrieve the content from the web. This new method yields richer QA data, including page titles and descriptions, and allows for a larger and more multilingual dataset. Additionally, we extend the scale of cross-lingual QA alignments, increasing their number from 1.5 million to over 13.8 million.

Second, we introduce a dataset of mined hard negatives to support the training of dense retrievers. This addition directly addresses feedback from the research community after the first WebFAQ release, where the absence of hard negatives was seen as a limiting factor for practical model training. The hard negatives were generated using a two-stage mining process: BM25 retrieval followed by reranking with BGE-m3~\cite{Chen2024-M3} as strong multilingual cross-encoder. This resource facilitates two training strategies: Contrastive Learning with MultipleNegativesRanking loss and Knowledge Distillation with MarginMSE loss. While the mined hard negatives improve training outcomes over random sampling in some configurations, we observe several important caveats. Most notably, false negatives remain a challenge, and random negatives still offer strong performance in contrastive setups. In contrast, Knowledge Distillation from the cross-encoder scores yields more robust improvements across non-English languages---but often at the cost of reduced performance in English.

Finally, WebFAQ 2.0 is part of a growing and continuous resource. Since November 2025, structured data including FAQs has been made available via the Open Web Index~\cite{Hendriksen2024}, with new dumps published daily.\footnote{\url{https://openwebindex.eu/owler/our_datasets}} This makes it possible to further expand WebFAQ in the future and to keep multilingual QA datasets up to date.

%The remainder of this paper is structured as follows: Section 2 reviews related work on multilingual QA and hard negative mining. Section 3 describes the revised data collection strategy. Section 4 introduces the hard negatives dataset. Section 5 presents experimental results and training strategies enabled by WebFAQ 2.0. Section 6 concludes with an outlook on ongoing work.

%% file: related_work.tex
\subsection{Multilingual and Cross-Lingual QA Datasets}

The first QA datasets expanding beyond English were primarily focused on reading comprehension~\cite{Clark2020, Lewis2020-mlqa, Artetxe2020-XQuAD}, while scalable retrieval benchmarks remained limited. Zhang et al.~\cite{Zhang2021} later addressed this shortcoming by offering human-annotated passage retrieval data across 11 languages. The MIRACL dataset~\cite{Zhang2023} further extended this to 18 languages, providing native speaker annotations for higher quality. These resources laid the foundation for evaluating multilingual retrievers.

To enable cross-lingual QA, Bonifactio et al.~\cite{Bonifacio2021-arxiv} translated English MS MAR\-CO queries and passages into other languages. Longpre et al.~\cite{Longpre2021} provided multilingual variants of Natural Questions paired with answers in over 25 languages. Thus, the two datasets adapted existing English QA datasets into multilingual test sets, enabling evaluation of cross-lingual understanding.

The original WebFAQ~\cite{Dinzinger2025} dataset contributed a large-scale resource of 96M QA pairs across 75 languages mined from schema.org FAQ markup. It also introduced monolingual retrieval benchmarks in 20 languages as well as 1.5M bilingual QA pairs, taking advantage of FAQ pages that exist in multiple languages. WebFAQ 2.0 expands upon this by both increasing data size and introducing hard negatives.

\subsection{Hard Negative Mining}

Effective training of dense retrievers often depends on the quality of negative examples. Hard negatives—passages that appear relevant but are not—are particularly useful. Qu et al.~\cite{Qu2021} introduced a method of denoising hard negatives using cross-encoders to remove potential false negatives. This significantly improved retrieval effectiveness on MS MARCO.

Cai et al.~\cite{Cai2022-arxiv} proposed Coupled Estimation Techniques to correct for pooling bias and reduce false negatives in training. Thakur et al.~\cite{Thakur2025} further explored this by using large language models (LLMs) to re-label training data, identifying mislabeled negatives and improving generalization.

Tirskikh et al.~\cite{Tirskikh2025} applied LLM-based filtering to explicitly judge the relevance of candidate negatives, reducing false negatives more effectively than score-based thresholds. Nguyen et al.~\cite{Nguyen2023} showed that even traditional lexical retrievers like BM25 can yield effective hard negatives when mined at the passage level rather than full-document granularity. This approach helps in identifying semantically misleading content and promotes more robust training signals.

Finally, Moreira et al.~\cite{Moreira2024-arxiv} proposed NV-Retriever, which leverages a novel variance-based method to identify valuable negative samples. Their approach assesses retrieval uncertainty to select informative negatives while reducing false negatives without relying on expensive cross-encoders.

%% file: data_collection.tex
\paragraph{Data Sources and Crawling}

Unlike the first version of WebFAQ, which used Web Data Commons (WDC) dumps as its data source, WebFAQ 2.0 crawls and extracts its own data. We started the crawl with a large list of URLs from the first dataset and augmented it with additional URLs mined from Common Crawl dumps from 2025. Specifically, we searched raw HTML content for the term \texttt{FAQPage}, indicating the presence of schema.org structured FAQ markup.\footnote{\url{https://schema.org/FAQPage}} These web pages were then downloaded directly using OWLer, a distributed web crawler developed as part of the OpenWebSearch project~\cite{Dinzinger2024}. This strategy yields three major advantages: (i) we obtain a significantly larger dataset than what WDC can provide, as WDC publishes structured data for only one Common Crawl snapshot per year; (ii) we extract \texttt{<link rel="alternate" hreflang="$\dots$">} references to multilingual versions of the same FAQ page, which improves our ability to mine bitext alignments; and (iii) we store additional metadata such as webpage titles and descriptions. This extra context helps to interpret QA pairs where the question is ambiguous in isolation.

\paragraph{Processing and Filtering}

The processing pipeline remains largely unchanged from the original WebFAQ~\cite{Dinzinger2025}. As before, we parse questions and answers from schema.org markup, clean HTML tags, and remove malformed entries. Language detection is performed using FastText~\cite{Joulin2017}. Previously, we used vector embeddings to identify and remove near-duplicate questions within the same site. The embeddings were computed using Jina v3~\cite{Sturua2024-arxiv}, a multilingual allrounder with good results on the MMTEB~\cite{Enevoldsen2025-openreview} Semantic Textual Similarity (STS) benchmark. In WebFAQ 2.0, we no longer apply this heuristic near-duplicate detection, as defining an appropriate threshold for semantic similarity is difficult. Instead, we sidestep this issue when building retrieval test sets by simply selecting only one QA pair per domain---ensuring test collections are unaffected by potential duplicates. The construction of retrieval test sets is still possible despite the limitation, simply due to the sheer number of mined FAQs.

\paragraph{Ambiguity and Contextual Relevance}

We still use Jina v3 embeddings to compute the semantic similarity score between each question and answer and release this score with the dataset, as it can help filter QA pairs where the question and answer are weakly related or require additional context.
This is necessary as some QA pairs are difficult to interpret without knowing the context of the page they originate from. For example, the question “Will it be painful?” provides no information about the subject, making it hard to interpret without additional information. The corresponding webpage title "Doubleview Podiatry – Walk with Confidence" suggests the topic is podiatric care. The answer "Most people experience little or no discomfort from the nail brace" clarifies the question’s meaning. By contrast, some QA pairs are clearly self-contained: e.g., “Is marijuana decriminalized in Colorado?” — “Yes, marijuana is decriminalized in Colorado.” The semantic similarity score between question and answer is helpful for identifying ambiguous pairs, such as the first example.

\begin{table}[h]
\centering
\caption{Top 10 Languages in WebFAQ 2.0 \textcolor{gray}{(v1 in gray)}}
\label{tab:top_langs}
\vspace{-0.2cm}
\begin{tabular}{p{1.8cm}rrp{0.01cm}p{2.1cm}rr}
\toprule
\textbf{Language} & \multicolumn{2}{c}{\textbf{\%}} & & \textbf{Language} & \multicolumn{2}{c}{\textbf{\%}} \\
\midrule
eng (English) & 27.9 & \textcolor{gray}{51.2} & & nld (Dutch) & 4.3 & \textcolor{gray}{2.8} \\
spa (Spanish) & 8.8 & \textcolor{gray}{6.0} & & por (Portuguese) & 4.0 & \textcolor{gray}{1.7} \\
deu (German) & 7.3 & \textcolor{gray}{6.9} & & ita (Italian) & 3.5 & \textcolor{gray}{2.7} \\
fra (French) & 6.5 & \textcolor{gray}{4.8} & & pol (Polish) & 2.5 & \textcolor{gray}{1.7} \\
rus (Russian) & 4.4 & \textcolor{gray}{3.8} &  & Other & 30.8 & \textcolor{gray}{13.0} \\
\bottomrule
\end{tabular}
\end{table}

\paragraph{Dataset Scale and Statistics}

WebFAQ 2.0 contains approximately \textbf{198 million QA pairs across 104 languages}, more than doubling the scale of the original release, which comprised 96 million QA pairs in 75 languages. While the first version already included a great share of non-English content, the language distribution has shifted even more significantly in this release. In particular, the share of English has decreased from over 50\% (49M) to under 30\% (55M), even though a language distribution with slightly over 50\% English content is common in multilingual resources mined from the web (e.g., see CCQA~\cite{Huber2022}). This shift in language distribution hence directly reflects our targeted efforts to expand multilingual coverage during data collection. Table~\ref{tab:top_langs} presents the top 10 languages in WebFAQ 2.0 by percentage of QA pairs, with the corresponding shares in the previous version shown in gray for comparison. Notably, the dataset now includes a broader and more balanced distribution across European, Slavic, and Romance languages, with a substantial increase in Dutch, Portuguese, and Polish content. But also many other languages have grown over-proportionally, such as Hindi from 537k to 2.6M, or Ukrainian from 806k to also 2.6M QA pairs.

Beyond increased language coverage, WebFAQ 2.0 also exhibits a notable shift in the distribution of topics. As shown in Table~\ref{tab:topics}, content related to Traveling and Hospitality now comprises nearly 60\% of the dataset—almost double its share from the previous release.
Following the approach from the original WebFAQ paper, topic labels were inferred using a fine-tuned XLM-RoBERTa classifier.\footnote{\url{https://huggingface.co/michaeldinzinger/xlm-roberta-base-qa-topic-classification}} For this version, we repeated the fine-tuning process with updated supervision: instead of using \texttt{GPT-4o-mini} as the labeling model, we employed \texttt{GPT-5-mini}, and extended each QA instance with additional context from the webpage title and description (when available) to support more accurate label generation. The classification model was fine-tuned for five epochs on a dataset of 79,650 \texttt{GPT-5-mini} annotated samples, drawn from the 49 most common languages (as in the previous version), ensuring one sample per website to promote topical and linguistic diversity. The data was split into 80\% training, 10\% validation, and 10\% test, resulting in a final F1 score of 88.07\% on the test set.

\begin{table}[t]
  \caption{Distribution of topics \textcolor{gray}{(v1 in gray)}}
  \label{tab:topics}
  \vspace{-0.2cm}
  \begin{tabular}{clrr}
  \toprule
  \multicolumn{2}{l}{\textbf{Topic}} & \multicolumn{2}{c}{\textbf{\%}} \\
  \midrule
  \faPlane & Traveling and Hospitality & 59.2 & \textcolor{gray}{34.1} \\
  \faShoppingCart & Products and Commercial Services & 18.9 & \textcolor{gray}{19.8} \\
  \faHeartbeat & Healthcare Services, Wellness and Lifestyle & 5.6 & \textcolor{gray}{13.0} \\
  \faMusic & Entertainment, Recreation, and Leisure & 4.7 & \textcolor{gray}{9.7} \\
  \faBank & Banking, Financial Services, and Insurance & 4.4 & \textcolor{gray}{6.0} \\
  \faInfoCircle & General Information and Other & 4.4 & \textcolor{gray}{3.9} \\
  \faGraduationCap & Employment, Education, and Training & 3.5 & \textcolor{gray}{9.5} \\
  \faGavel & Legal Services, Regulations and Government & 2.3 & \textcolor{gray}{4.0} \\
  \bottomrule
  \end{tabular}
\end{table}

\begin{table}[h]
\centering
\caption{Distribution of Question Types in WebFAQ 2.0}
\label{tab:question_types}
\begin{tabular}{lrr}
\toprule
Question Type & \multicolumn{1}{c}{Count} & \multicolumn{1}{c}{\textbf{\%}} \\ 
\midrule
Not-a-Question & 67{,}870{,}478 & 34{,}7 \\
Factoid & 65{,}601{,}236 & 33{,}6 \\
Experience & 22{,}649{,}722 & 11{,}6 \\
Instruction & 20{,}179{,}307 & 10{,}3 \\
Evidence-Based & 8{,}300{,}537 & 4{,}3 \\
Reason & 6{,}503{,}638 & 3{,}3 \\
Comparison & 4{,}236{,}223 & 2{,}2 \\
Debate & 107{,}626 & 0{,}1 \\
\bottomrule
\end{tabular}
\end{table}

\paragraph{Question Types.}

We also categorize QA pairs by question type using the taxonomy proposed by Bolotova et al.~\cite{Bolotova2022}. Their work defined seven non-factoid question types (e.g., Instruction, Comparison, Reason, etc.) for English, but we extend this to a multilingual setting. We trained an XLM-RoBERTa model on a new multilingual dataset, spanning 49 languages and comprising about 48,300 labeled question instances.\footnote{\url{https://huggingface.co/datasets/AliSalman29/nfqa-multilingual-dataset}} The category labels were not human-annotated but generated via an ensemble of three large language models (Meta’s LLaMA 3.1~\cite{Meta2024-arxiv}, Google’s Gemma 2~\cite{Gemma2024-arxiv}, and Alibaba’s Qwen 2.5~\cite{Qwen2024-arxiv}) that independently classified each question and assigned a final label by majority vote.
Additional questions were synthetically created to ensure each language–category pair had sufficient examples, and these were re-annotated by the same LLM ensemble for consistency. Again, we fine-tuned an XLM-RoBERTa model on this data, for 6 epochs with learning rate $2 \times 10^{-5}$ and batch size 16.\footnote{\url{https://huggingface.co/AliSalman29/nfqa-multilingual-classifier}} The resulting classifier achieved approximately 88.1\% F1 score on the test set. However, because these labels were obtained automatically via LLM predictions rather than manual annotation, this multilingual classification step is regarded as preliminary and in need of further refinement. Nevertheless, we have included the obtained question type labels in the dataset (see Table~\ref{tab:question_types}). Analogously to the topic labels, computed for the 49 most common languages.

\paragraph{Bilingual QA Alignments}

Bitext mining is important for training and evaluating cross-lingual retrieval models and multilingual sentence embeddings. In WebFAQ 2.0, we mine bilingual QA alignments following the same general approach as in the original release~\cite{Dinzinger2025}. Specifically, we embed all questions and answers across language pairs using LaBSE (Language-Agnostic BERT Sentence Embeddings)~\cite{Feng2022} and retrieve nearest neighbors to form aligned question–answer pairs. 
% Before bitext mining, we apply a minimum semantic similarity threshold of 0.5 to discard poor-quality QA pairs.
To ensure high alignment quality, we apply a strict minimum LaBSE similarity threshold of 0.9. Using this criterion, we identify aligned QA pairs for 3,970 language combinations that each contain at least 100 aligned samples. Among those, 1,282 language pairs contain at least 4,000 aligned samples.

In WebFAQ v1, we released language combinations with at least 100 aligned samples, resulting in 1,001 language pairs and 1.5M aligned QA pairs in total. In contrast, WebFAQ 2.0 increases the total number of aligned QA pairs to over 14.3M across all 3,970 language pairs. Even when restricting the count to the 1,282 combinations with at least 4,000 samples, the dataset still contains more than 9.6M aligned QA pairs. This presents a substantial expansion in both scale and language coverage.\footnote{\url{https://huggingface.co/datasets/michaeldinzinger/webfaq-v2-bitexts}}

Table~\ref{tab:bitext} illustrates the increase in alignment counts for the top five language pairs in v2. Notably, the strongest growth is not limited to the usual high-resource pairs involving English. Numerous alignments emerge between non-English language pairs, including combinations beyond European languages. The largest alignment count is observed for Marathi–Telugu, two major regional languages in India, demonstrating that the expansion of WebFAQ 2.0 meaningfully improves coverage for underrepresented language pairs.

\begin{table}[h]
\centering
\caption{Top language pairs for bilingual QA alignment.}
\label{tab:bitext}
\begin{tabular}{lrcr}
\toprule
Language Pair & \multicolumn{1}{c}{\textbf{v1}} &  & \multicolumn{1}{c}{\textbf{v2}} \\ \midrule
Marathi–Telugu & 0 & $\rightarrow$ & 89,910 \\
German–Spanish & 19,739 & $\rightarrow$ & 80,623 \\
Russian–Ukrainian & 15,251 & $\rightarrow$ & 74,545 \\
Italian–Portuguese & 10,924 & $\rightarrow$ & 69,469 \\
Indonesian–Korean & 1,515 & $\rightarrow$ & 42,986 \\
\bottomrule
\end{tabular}
\end{table}

To validate alignment quality, we used the GEMBA metric~\cite{Kocmi2023} based on LLM judgments. This metric confirmed the alignment accuracy for a sample of mined pairs. We also contributed a new bitext mining task to the MTEB~\cite{Muennighoff2023} benchmark,\footnote{\url{https://github.com/embeddings-benchmark/mteb/blob/main/mteb/tasks/bitext_mining/multilingual/web_faq_bitext_mining.py}} complementing the retrieval task submitted as part of WebFAQ v1.\footnote{\url{https://github.com/embeddings-benchmark/mteb/blob/main/mteb/tasks/retrieval/multilingual/web_faq_retrieval.py}} The exact methodology and evaluation of the bitext mining approach is reported by Dinzinger et al.~\cite{Dinzinger2025} and demonstrates excellent translation quality.
% Together, these updates make WebFAQ 2.0 a scalable and extensible dataset for training and evaluating multilingual retrieval and bitext mining systems.

%% file: hn_collection.tex
To support more effective training of dense retrieval models, we release a hard negatives dataset derived from the WebFAQ corpus. We adopt a two-stage mining process guided by previous research~\cite{Qu2021}, designed to yield challenging but non-relevant candidate answers while mitigating the risk of false negatives.

In the first stage, we use BM25 to retrieve the top 200 candidate answers for each query from the corresponding WebFAQ monolingual corpus. These candidates typically share lexical overlap with the question and represent plausible yet incorrect responses. In the second stage, we rerank these 200 candidates using BGE-m3~\cite{Chen2024-M3}, a strong cross-encoder model.
The cross-encoder relevance scores are meant to be utilized, among others, for \textit{denoising}, as described in RocketQA~\cite{Qu2021}, where candidates are filtered based on an adaptive threshold, to reduce the risk of false negatives.
% To ensure correctness, we filter these candidates by checking whether their cross-encoder relevance scores fall below an adaptive threshold—thereby reducing the risk of false negatives, similar to the denoising process described in RocketQA~\cite{Qu2021}.

The final output is a hard negatives dataset consisting of quintuples, formatted as (query, positive, pos\_score, negatives, neg\_scores). The dataset includes exactly 1.25 million such quintuples across 20 languages, with the smallest language splits, such as Danish or Vietnamese, comprising 32k samples. We provide this dataset alongside the exact language distribution publicly via HuggingFace.\footnote{\url{https://huggingface.co/datasets/IrvinTopi/WebFAQHardNegatives}}

Our dataset enables two prominent training strategies:
\begin{itemize}
\item \textbf{Contrastive Learning:} With MultipleNegativesRanking\-Loss (MNR)~\cite{Henderson2017-arxiv}, paired triplets are used to contrast positives and hard negatives. The cross-encoder scores allow us to apply denoising strategies and explicitly select the negatives per query.
\item \textbf{Knowledge Distillation:} Using the cross-encoder scores directly in the loss function via MarginMSE~\cite{Hofstaetter2020-arxiv}, the dense retriever learns to mimic the fine-grained judgments of the strong multilingual reranker.
\end{itemize}

% In Section 5, we demonstrate how both approaches benefit from the inclusion of these carefully mined negatives.

% \textcolor{blue}{The final output is a dataset of hard negatives derived from the top 200 reranked candidates. Each negative is difficult to distinguish from the positive in terms of content but has been filtered to avoid semantic correctness. The dataset includes approximately 1.3 million samples across 20 languages. We provide this dataset publicly via HuggingFace.}

% \textcolor{blue}{Our dataset enables two prominent training strategies:}
% \begin{itemize}
%     \item \textcolor{blue}{\textbf{Contrastive Learning:} We prepare a specific training format with a \textbf{1:4 ratio} (one positive, four hard negatives). With MultipleNegativesRankingLoss (MNRL), these randomly paired triplets utilize the filtered negatives to contrast positives against challenging non-relevant candidates while minimizing label noise.}
%     \item \textcolor{blue}{\textbf{Knowledge Distillation:} We utilize the re-ranker output, including the cross-encoder scores. Using MarginMSE~\cite{Hofstaetter2020-arxiv}, the dense retriever learns to mimic the fine-grained judgments of the reranker directly from the score distribution.}
% \end{itemize}

%% file: evaluation.tex
\begin{table*}[t]
\caption{Comparing retrieval performance on 3 multilingual datasets using NDCG@10 in \%, including BM25 and XLM-RoBERTa-base with in-domain pre-training on MS MARCO (\textit{Base}) as references. \textit{MNR (RN)} represent the \textit{Base} model fine-tuned on WebFAQ random negatives. The three bottom rows represent training with hard negatives. The top scores are marked in bold, while the second-best are underlined. Using the reranker scores directly in the loss function yields the best results (\textit{M-MSE}).}
\label{tab:results}
\vspace{-0.2cm}
\addtolength{\tabcolsep}{-0.07em}
\begin{tabular}{rcccccccccccccccccccc}
 & \multicolumn{6}{c}{WebFAQ} &  & \multicolumn{6}{c}{MIRACL (Hard Negatives)} &  & \multicolumn{6}{c}{Mr. Tydi} \\ \cline{2-7} \cline{9-14} \cline{16-21} 
 &  &  &  &  &  &  &  &  &  &  &  &  &  &  &  &  &  &  &  &  \\[-10pt] 
 & ara & eng & ind & jpn & kor & rus &  & ara & eng & ind & jpn & kor & rus &  & ara & eng & ind & jpn & kor & rus \\ \midrule[1.5pt] % \cline{2-21} 
 &  &  &  &  &  &  &  &  &  &  &  &  &  &  &  &  &  &  &  &  \\[-10pt]
BM25 & 29.7 & 24.4 & 36.1 & 29.0 & 29.6 & 21.2 &  & \underline{53.2} & 31.6 & \textbf{51.6} & \underline{44.7} & 37.6 & 29.8 &  & 43.4 & 19.9 & \underline{50.5} & 25.0 & 24.8 & 29.7 \\ \midrule
Base & 61.0 & 49.7 & 74.6 & 56.5 & 73.2 & 50.1 &  & 36.1 & 30.6 & 28.1 & 27.6 & 39.3 & 29.6 &  & 30.5 & 18.1 & 33.2 & 16.0 & 27.7 & 22.5 \\ \midrule[1.5pt]
MNR (RN) & 74.3 & \textbf{60.0} & \underline{81.9} & \textbf{69.8} & \underline{81.3} & \textbf{62.0} &  & 50.5 & \textbf{37.2} & 36.1 & 39.8 & \underline{42.4} & \underline{39.6} &  & \underline{44.7} & \textbf{29.5} & 47.7 & \underline{33.5} & \underline{34.7} & \underline{30.8} \\ \midrule
MNR Top4 & 71.8 & 48.7 & 78.8 & 63.6 & 81.4 & 51.4 &  & 40.3 & 23.8 & 26.6 & 31.4 & 35.3 & 27.6 &  & 34.1 & 13.0 & 36.9 & 25.1 & 26.9 & 19.2 \\
MNR Den. & \underline{74.8} & \underline{57.8} & \textbf{82.2} & 69.6 & 83.1 & 59.9 &  & 46.4 & 29.7 & 31.7 & 36.6 & 39.0 & 34.8 &  & 39.5 & 20.1 & 44.7 & 30.1 & 32.6 & 26.9 \\ \midrule
M-MSE & \textbf{75.4} & 57.4 & 78.5 & \underline{69.7} & \textbf{81.5} & \underline{61.1} &  & \textbf{54.7} & \underline{34.7} & \textbf{38.7} & \textbf{46.0} & \textbf{50.9} & \textbf{42.6} &  & \textbf{50.4} & \underline{27.7} & \textbf{51.4} & \textbf{35.6} & \textbf{37.6} & \textbf{40.3} \\ \bottomrule[1.5pt]
\end{tabular}
\end{table*}

We evaluate the effect of training with mined hard negatives on the performance of multilingual dense retrievers. Our goal is to show that fine-tuning with WebFAQ 2.0’s hard negatives leads to more effective models than training with randomly sampled negatives.

\subsection{Setup}

\paragraph{Base model}

We use an XLM-RoBERTa base encoder that has been in-domain pretrained on MS MARCO using MarginMSE loss \cite{Hofstaetter2020-arxiv}, as outlined in the original WebFAQ paper~\cite{Dinzinger2025}. This pretrained model serves as a strong multilingual baseline.\footnote{\url{https://huggingface.co/PaDaS-Lab/xlm-roberta-base-msmarco}}

\paragraph{Fine-tuning configuration}

All models are fine-tuned on 1.25 million training samples covering 20 languages. Following established practices in dense retrieval~\cite{Qu2021}, we employ a negative sampling ratio of 1:4 (one positive to four hard negatives per query) within the Contrastive Learning setup. We experiment with two hard negatives training strategies, where the first configuration, using random negatives, serves as reference. To ensure fair comparison, we train all configurations for one epoch with a batch size of 128.

\begin{itemize}
\item \textbf{Random negatives (RN):} Triplets are formed using randomly sampled negatives.
\item \textbf{Hard negatives (MNR):} Triplets include our mined hard negatives, fine-tuned through Contrastive Learning with MultipleNegativesRankingLoss. We apply two strategies of selecting the hard negatives: (1) \textit{Top4}, which selects the top 4 results from the cross-encoder reranked candidates; and (2) \textit{Denoised}, which follows RocketQA~\cite{Qu2021} by discarding negatives with cross-encoder scores greater than 0.9 or less than 0.1 and randomly sampling 4 from the remaining set.
\item \textbf{Hard negatives (M-MSE):} The same hard negatives are used, but fine-tuned using MarginMSE loss to incorporate the cross-encoder scores directly as soft targets. Per query, the top 10 ranked hard negatives are used for training.
\end{itemize}

\paragraph{Evaluation benchmarks}

We evaluate model performance on three multilingual retrieval benchmarks. The first is WebFAQ Retrieval, which covers 20 languages and serves as an in-domain evaluation setting. The second is MIRACL-HN~\cite{Zhang2023}, a multilingual benchmark specifically constructed with hard negatives to assess retrieval robustness. The third is Mr. TyDi~\cite{Zhang2021}, which tests zero-shot multilingual passage retrieval across diverse languages. For each benchmark, we report NDCG@10 results across six representative languages: Arabic, English, Indonesian, Japanese, Korean, and Russian.

\subsection{Results}

Table~\ref{tab:results} compares the performance of the base model, fine-tuned with random negatives (RN), to models fine-tuned with hard negatives using contrastive learning (MNR) and knowledge distillation (M-MSE). While our resource enables two common training paradigms, the experiments reveal important caveats in this regard. First, we observe that \textbf{false negatives are prevalent} in our mined hard negatives. As a consequence, models trained via Contrastive Learning using random negatives (RN) perform better overall than those trained with unfiltered hard negatives. We see that this gap is more pronounced for \textit{Top4} than for \textit{Denoised}, when hard negatives are randomly sampled and filtered to exclude those very similar to the query. Hence, applying denoising yields improvements under the MNR setting, but RN remains more robust.

In contrast, \textbf{Knowledge Distillation via MarginMSE (M-MSE)} benefits from incorporating the ranking scores of the cross-encoder directly into the loss function. This training yields a stronger model that outperforms the MNR configurations across most non-English languages. Nonetheless, this advantage does not extend to English. Since the base model was in-domain pretrained on MS MARCO (exclusively English) and the hard negatives also include a large share of English training data, we observe a performance trade-off: gains in non-English retrieval come at the cost of reduced performance in English.

These findings highlight that while WebFAQ 2.0’s hard negatives enable advanced training strategies, effective application still requires careful handling---particularly with regard to false negatives and the generalization of performance gains across languages.

% The results show that both models trained on hard negatives (MNRL and M-MSE) outperform the random negative baseline (RN) across all benchmarks. The MNRL model provides strong improvements via contrastive learning, while the M-MSE model demonstrates that the same dataset can support knowledge distillation by leveraging the cross-encoder scores.

% This confirms the utility of the WebFAQ 2.0 hard negatives dataset, which enables multiple training strategies and improves multilingual retrieval quality both in-domain and in zero-shot settings.

% \textcolor{blue}{The results demonstrate the high utility of the WebFAQ 2.0 hard negatives dataset for advanced training strategies. While standard contrastive learning (MNRL) yields competitive in-domain performance, the true value of the mined negatives is unlocked through knowledge distillation. The M-MSE model, which leverages the fine-grained scores provided in our dataset, consistently outperforms the strong random negative baseline across all benchmarks. On the challenging out-of-domain MIRACL dataset, using our hard negatives leads to significant gains (e.g., Arabic: 54.7 vs 50.5; Korean: 50.9 vs 42.4). This confirms that the proposed dataset provides the necessary signal to robustly improve multilingual retrieval quality and zero-shot generalization.}

%% file: conclusion.tex
We presented WebFAQ 2.0, an expanded multilingual QA dataset and a new hard negatives resource to support training and evaluation of dense retrieval models. Our work makes two key contributions: a significantly enlarged and more diverse collection of QA pairs with extended bilingual coverage, and a carefully curated hard negatives dataset created through a two-stage mining process. Together, these resources enable more effective training strategies and highlight the trade-offs involved in fine-tuning dense retrievers, especially in multilingual settings.

Looking ahead, the continuous publication of FAQs and structured data as part of the Open Web Index~\cite{Hendriksen2024} offers a clear path for growing WebFAQ over time. We plan to explore this new data source to build even more extensive and timely multilingual QA benchmarks in future work.